# Reprogrammable holograms from maskless surface photo-morphing

**Francesco Reda,[1] Marcella Salvatore,[1,2] Marco Astarita,[3] Fabio Borbone[4] and Stefano L. Oscurato[1,2,*]**

[1] *Physics Department "E. Pancini", University of Naples "Federico II", Complesso Universitario di Monte Sant'Angelo, via Cinthia 21, 80126, Naples, Italy*

[2] *Centro Servizi Metrologici e tecnologici Avanzati (CeSMA), University of Naples "Federico II", Complesso Universitario di Monte Sant'Angelo, Via Cintia 21, 80126, Naples, Italy*

[3] *Physics Department, Politecnico di Milano, 20133, Milan, Italy*

[4] *Department of Chemical Sciences, University of Naples "Federico II", Complesso Universitario di Monte Sant'Angelo, Via Cintia, 80126 Naples, Italy*

*\*Corresponding author: stefanoluigi.oscurato@unina.it*

## Abstract

Holographic technologies have the potentiality to impact our everyday life in many sectors including science, education, entertainment, art, and healthcare. Although holographic screens and projectors are part of common imagination since long time, they are still at initial stages of development and integration. Recent achievements of metasurface and flat optics research gave an unprecedented strength to this field, overcoming critical aspects as efficiency, size and flexibility of conventional optics and liquid crystal technologies. However, although diffractive and metasurface holographic projectors with advanced functionalities and improved efficiencies are continuously reported, they are static devices, requiring demanding, burdensome, and irreversible manufacturing processes. Here we report an all-optical and single-step lithographic framework for the fabrication of diffractive holographic projectors directly on the surface of a photo-morphable polymer film. Real-time optimization during the accurate surface patterning and fully structural reconfigurability allowed for the first prototype of a fully reprogrammable pixel-less morphological projector, opening new routes for holographic image displaying and optical data sharing.

# 1. Introduction

Light-modulating planar devices can empower many emerging technologies as virtual and augmented reality (*1–3*), optical wireless communication (*4, 5*), green energy harvesting (*6, 7*), opening also to the next-generation of displays and holographic projectors (*8*). Despite holograms can be implemented through addressable liquid crystals on silicon (LCOS) devices (*9–12*), diffractive optical elements (*13–15*) and metasurfaces (*16–18*) are increasingly gaining interest for holographic applications, due to their ability to generate arbitrary optical fields from the modulation of an incident light beam through a ultra-compact and planar device. Untied from electronics, efficiency, and size limitations of LCOS displays, planar holographic devices promise a greater possibility of miniaturization while maintaining higher efficiencies and light modulation capabilities. In addition to images projection, planar holographic devices can also represent a valid platform for optical information storing, encryption and sharing (*19–21*). Nevertheless, as also valid for holographic displays, those technologies intrinsically require optical supports able to be fully erased and rewritten, a milestone only partially achieved with several limitations by tunable metasurfaces (*22–24*). However, these features come at the expense of realization of complex surface geometries at light (sub-)wavelength scale, where the manufacturing process, typically leading to static devices, (*25, 26*) can pose severe performance and/or economic limitations.

Optical lithography is among the most used surface patterning techniques (*27*) for the fabrication of planar optical devices. Starting with the irradiation of a photoresist by means of a structured illumination pattern produced by a mask, the typical workflow requires additional post-exposure chemical, physical and mechanical processes, through which the desired surface pattern is finally transferred to the operating device (*27, 28*). The multileveled surface patterns needed for optimal functionality of a holographic device can even require several iterations of this scheme (*13*). Maskless methods, where the multistep mask exposure is replaced digital-based projection of spatially structured intensity patterns over the photoresist surface, can offer however greater control and flexibility for the realization of the complex lateral geometry and the grayscale modulation (*29*). Both deformable micromirror devices (DMDs) (*30, 31*) and LCOS (*32–34*) were explored as programmable spatial light modulators to achieve digital maskless surface patterning for optical devices manufacturing. In addition, even non-optical maskless approaches as particle beam fabrication methods (*35, 36*) and scanning probe lithography (*37, 38*) have been reported for the accurate optical device manufacturing, but these methods suffers of reduced throughput, increased costs and energetic impact with respect to optical techniques (*26, 27*).

Here, we demonstrate the direct all-optical maskless fabrication of fully reconfigurable diffractive holographic devices, implemented as thin structured transmissive phase retarders realized on the surface of a reprogrammable dielectric material. To this aim, a digital holographic optical scheme is used to generate and project a grayscale spatially structured intensity distribution of light on an azobenzene-containing polymer film, whose surface locally deforms according to the irradiated spatial light distribution. In this way, the structured surface of the operating optical device is directly produced without any additional lithographic step. The photomechanical process responsible for the direct surface morphing of the polymer is intrinsically reversible, allowing the update of the fabricated surface geometry at will. Compared to other optical maskless techniques, our approach allows to fully exploit the possibility of arbitrary spatiotemporal modulation of the holographic writing beam and the integration of the lithographic system with a real-time optical characterization setup allowing to evaluate the device performance already during the fabrication. The all-optical system is used here to realize operating optical configurations and devices with advanced and optimized functionalities, including reprogrammable grayscale holograms with improved visibility and tunable axial position, and a high-density optical encryption scheme able to temporally split secreted holographic information. Representing a new state-of-the-art as a reprogrammable all-optical fabrication framework for custom multileveled flat optical devices, our approach can assist the development of next generation photonics, starting from devices prototyping, testing and assembly until to their large-scale distribution.

## 2. Results
### 2.1. Direct holographic surface structuration

To elucidate the main features of our direct holographic maskless surface patterning scheme, schematically represented in Fig. 1A, we first demonstrate the realization of simple arbitrary binary pattern on the surface of an azobenzene-containing polymer thin film (herein referred to as *azo-resist* to highlight its functionality as lithographic material). This class of amorphous materials exhibit the unique property of stable surface reliefs formation under low-intensity structured UV-visible light irradiation (*39*, *40*) as a consequence of a directional material transport initiated by the azobenzene chromophores hosted in the polymeric matrix (*41–43*), with a mechanism still to be fully unveiled (*44*) . Due to the sensitivity to both the intensity and the polarization of the irradiated light, the surface reliefs on azopolymer films enable a direct vectorial lithography, exploited in many configurations, including interference, high-fusing, near-field, and pure structured polarization illumination (*45–54*) .

In the irradiation of a circularly polarized light patten $I_W(x, y)$ in a low-focusing regime, the spatiotemporal evolution of the surface morphology $h(x, y, t)$ can be phenomenologically described as (*54*, *55*):

$$h(x, y, t) = \nabla^2[I_W(x, y)] \cdot h_0(t) \qquad (1)$$

In a low intensity regime, the relief depth $h_0$ increases approximately linearly with the exposure time ($h_0(t) = c \cdot t$, where $c$ is a phenomenological inscription efficiency constant), while the material flows from the high intensity region toward dark areas (as schematized in Fig.

1B), forming then a surface relief pattern with the same geometry as the illuminating intensity $I_W(x,y)$.

In our maskless optical lithographic scheme, we used a phase-only Computer-Generated Hologram (CGH) system to fully exploit the direct surface structuration process described by eq.(1). In this configuration (see also Materials and Methods), arbitrary grayscale illumination patterns $I_W(x,y)$, originated by a computer-controlled phase-only Spatial Light Modulator (SLM), can be directly transferred to the entire illuminated area of the polymer surface in a single exposure step.

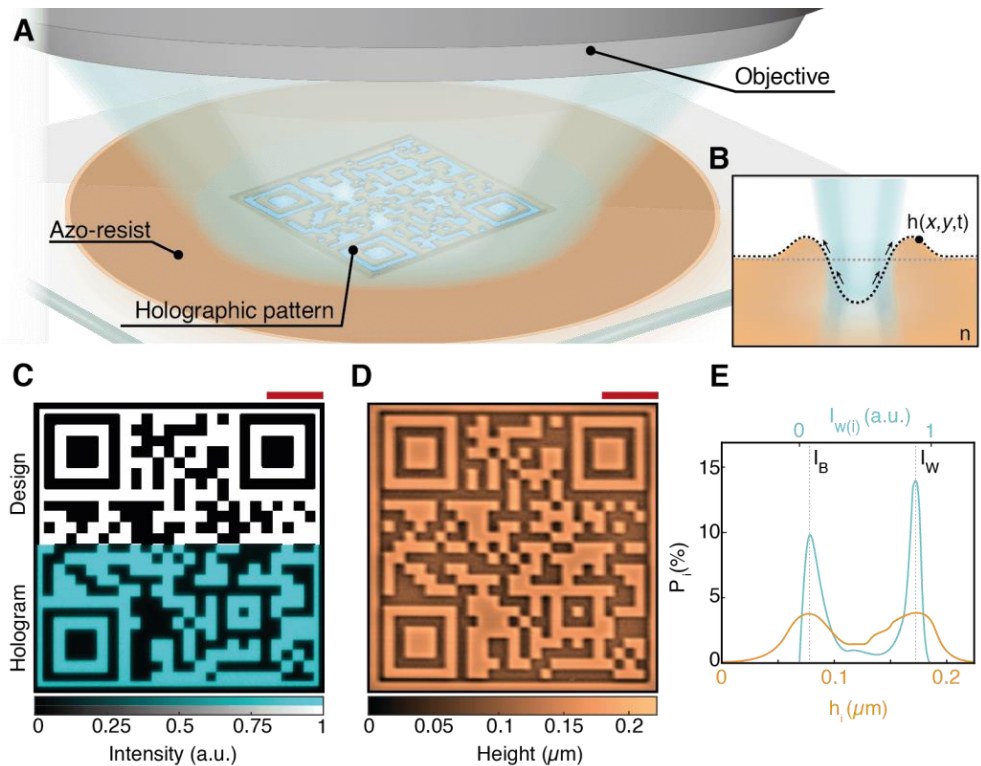

Fig. 1. Holographic structuration of azo-resist surface. **A** Graphical representation of the holographic inscription scheme. Writing beam, with arbitrary shaped intensity profile is directly projected over the azo-resist surface by an objective. **B** Light triggered mass migration occurring at surface of amorphous azopolymer films under structured illumination absorption, leading to stable surface geometries $h(x,y,t)$. **C** Design and reconstruction of a QR code shaped holographic pattern. The experimental intensity pattern is the result of time averaging of the holographic sequence, allowing to reduce speckle noise effects. **D** Atomic force microscope micrograph of the structured surface collected right after the exposure step. Red scale bar, both in panels C and D corresponds to a physical size of $20\ \mu m$ on the sample. **E** Height distribution probability (orange plot) compared with the intensity probability distribution (sky blue plot) of the holographic beam. Each point of the line plot represents the probability $P_i$ of having a fixed height value $h_i$ in the AFM image corresponding to the implemented intensity level $I_{(w)i}$.

To demonstrate our ability in arbitrary direct surface patterning, we designed a two-levels QR code as an 8-bit two-dimensional image, from which the illuminating light pattern $I_W(x,y)$ is calculated (*56*) (Fig. 1C top). The generated holographic writing pattern is projected on the surface of the azo-resist by means of a long-working distance microscope objective, where a relief pattern $h(x, y)$ directly appears. Additional details about the writing holographic design, the illumination homogeneity improvement, and the resolution of our configuration can be found in Methods section and in Fig. S4.

Fig. 1D shows the Atomic Force Microscope (AFM) micrograph of the polymer film surface after being exposed to the holographic pattern for $t = 20\ s$. AFM image is collected right after the exposure step without any additional post-exposure process. The surface relief pattern faithfully reproduces the target image, and, as expected from eq. (1), is the complementary of the illuminating hologram.

To extend the visual comparison to a quantitative analysis needed in the fabrication of complex relief pattern from the design of a diffractive phase-modulating mask acting as holographic projector, we characterize eventual mismatch errors between the target and the experimental surface morphology described in Fig. 1. To this aim, we retrieved the height distribution of the surface form the topographic image, with a sampling interval of $0.387\ \mu m$ determined by the pixel size of the AFM scan (see also Methods). The distribution, shown in Fig. 1E, must be compared with the target one, in which there are only two equally weighted levels corresponding to the black ($I_B$) and white ($I_W$) pixels of the image. Despite the presence of two narrow bands in the distribution extracted from the optical image of the hologram (blue curve in Fig. 1E), confirming the high contrast in the writing binary pattern, the topographic distribution (orange curve) of the two heigh levels appears broadened. The origin of such structural mismatch resides in the relief smoothing at illumination edges with sharp contrast jumps (see also Fig. S5), as predicted for the light-induced material transport phenomenology described by eq. (1). In our previous implementation of this lithographic method, we circumvented this issue by limiting quantitative design to smooth sinusoidal surfaces (*46, 57*). However, sharp features could potentially be encoded in the design of a suitable optimized holographic pattern associated to the target image, providing eventually a narrower topographical distribution when transferred on the azo-resist film. As further detailed below, the all-optical scheme used here to fabricate and simultaneously characterize the diffractive optical components allows the minimization of the effects on optical performances originated by similar fabrication-design mismatches inherent to the simplified description of material transport in hologram design.

Even in the simplistic linear response relief design used here, the results in Fig. 1 fully demonstrate the potentialities of our scheme as a direct maskless holographic technique for the arbitrary structuration of the surfaces at the microscale. The fidelity of the surface pattern can be further demonstrated by to the possibility of effectively read the binary QR code (by any camera QR code reading software) from the topographic data, rendered as two-dimensional image with a linear colormap (Fig. 1D).

## 2.2. Holographic morphological projectors: design, optimization, and fabrication

For the design of the azopolymer-based morphological holographic projectors we leverage the results of the scalar diffraction theory (*10*). While conventional projection displays exploit amplitude-modulating pixels to locally and selectively block part of the incident light to form images, a diffractive holographic projector can be implemented as a phase-only planar device for a coherent monochromatic light modulation (*10*, *12*), able to reconstruct a desired light pattern without making use of absorption phenomena. Phase-only holographic plates, named *kinoforms* (*58*), can implement the proper modulating complex transmission function $t(x,y) = \exp(i\varphi(x,y))$ as local thickness variations $h(x,y)$ of a dielectric material (Fig. 2), which influence the optical path traveled by an input monochromatic field $U_{in}(x,y,z_{in})$ (see Materials and Methods). The phase mask $\varphi(x,y)$ is typically referred to as *kinoform* (*58*).

According to the diffraction theory, in the case of far-field propagation $z \gg z_{in}$ (Fraunhofer approximation, where), the emerging modulated field $U_{out}(x,y,z)$ is two-dimensional spatial Fourier transform of the beam modulated at the kinoform plane, resulting in a reconstructed image $I_{out}(x,y)$ determined by the relation (*10*):

$$I_{out}(x,y,z) = \left|FT\left[U_{in}(x,y,0) \cdot e^{i\varphi(x,y)}\right]\right|^2 \qquad (2)$$

An analogous result can be also found between the two focal planes of a thin lens, reducing the image reconstruction to finite distances (*10*). By inversion of eq. (2), the kinoform $\varphi(x,y)$, and the relative mask surface relief pattern $h(x,y)$ for any given target holographic image $I_{out}(x,y)$ could be potentially calculated. However, for a phase-only modulator, the kinoform can be retrieved only through iterative algorithms (*8*). Fig. 2 schematically shows this process for the case for the desired output image $I_{out}$ representing the Greek letter "π", where the conventional Gerchberg–Saxton (GS) algorithm (*59*) is used as iterative Fourier transform algorithm (IFTA) to retrieve the kinoform $\varphi(x,y)$.

Once the kinoform is calculated, all the challenges involved in the fabrication of the holographic projector are shifted to manufacturing level. Optimal image reconstruction requires an accurate transfer of the designed phase mask, including the position of the phase discontinuities (lateral pattern) and the value of local and maximum phase delays, in the proper surface relief pattern. Any defect arising in this process deteriorates the hologram quality, causing the reduction in the diffraction efficiency and the appearance of spurious contributions in the target holographic image, consisting of an unmodulated optical component (DC term) and several shifted and scaled replicas of the desired intensity pattern (ghost or false images) (*60*). These contributions can overlap in the reconstruction plane, requiring eventually an off-axis design for the hologram (Fig. 2), which reduces the available target image domain by half of the field of view (*61*). However, even in the case of a defect-free lateral pattern transfer, a deviation from a full $2\pi$ modulation depth, associated with eventual total relief heigh errors induced in the dielectric structured surface, still cause the emergence of the spurious holographic terms. To reduce this effect, an ideal optimal modulation depth of $h_0 = \lambda/(n-1)$ should be realized. This condition simultaneously grants the maximization of the diffraction efficiency in the target holographic image and ghost hologram suppression (see also Supplementary Information).

In our direct lithographic scheme, the surface relief pattern $h(x, y)$ and the modulation depth $h_0$ can be independently controlled by the digital holographic design and by the exposure time, respectively. Then, the generalization of the inscription scheme of Fig. 1 to the projection of a grayscale structured light pattern with the geometry of a calculated kinoform $\varphi(x, y)$ can lead to the fabrication of optimized morphological holographic projectors directly as a surface relief pattern on the dielectric azo-resist film.

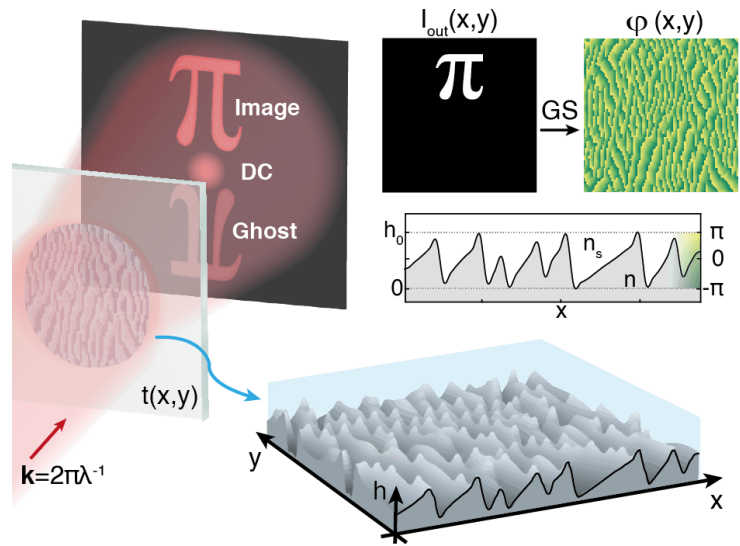

Fig. 2. Design of holographic morphological projectors. Target intensity $I_{out}$ is used to retrieve, by GS iterative algorithm, the proper phase map $\varphi(x, y)$ to be implemented as dielectric height modulated phase retarder. The material with refractive index $n$ is assumed to be immersed in a surrounding medium with refractive index $n_s$. When illuminated with monochromatic light, with wavevector $k = 2\pi/\lambda$, the phase retarder (kinoform) produces a diffracted beam depending on the optical delay accumulated by the light passing through the structured surface. The kinoform allows the reconstruction of the target holographic image defined during the design and additional spurious diffraction orders to be suppressed by tuning the total modulation depth $h_0$.

To this aim, we first characterized the ability of our lithographic scheme in encoding multiple discrete intensity levels of light in a single holographic pattern, useful to calibrate the response our system for the generation of the complex grayscale pattern required by a kinoform fabrication (see Fig. S6).

Then, we directly inscribe on the azopolymer film the grayscale surface profile $h(x, y)$ of the kinoform calculated for the reconstruction of far-field holographic image of the Greek letter "π".

In this process, the 8-bits (256 levels) digitally calculated kinoform $\varphi(x,y)$ is converted into a gray-scale holographic pattern $I_W(x,y)$, which induces the correspondent relief pattern $h(x,y)$ on the azo-resist surface (Fig. 3). For the analysis of the lateral pattern and the determination of the total height excursion $h_0$ of the produced surface relief, we performed SEM and AFM analysis after the exposure process. The SEM analysis (Fig. 3) confirms a correct position-matching of the phase discontinuity in the kinoform, granting a global correct relief lateral geometry. Fig. 3A shows, instead, the three-dimensional topographic micrograph of a portion of a typical azo-resist kinoform surface, evidencing the continuous heigh variation in the pattern, encoded in the grayscale writing holographic pattern (Fig. 3).

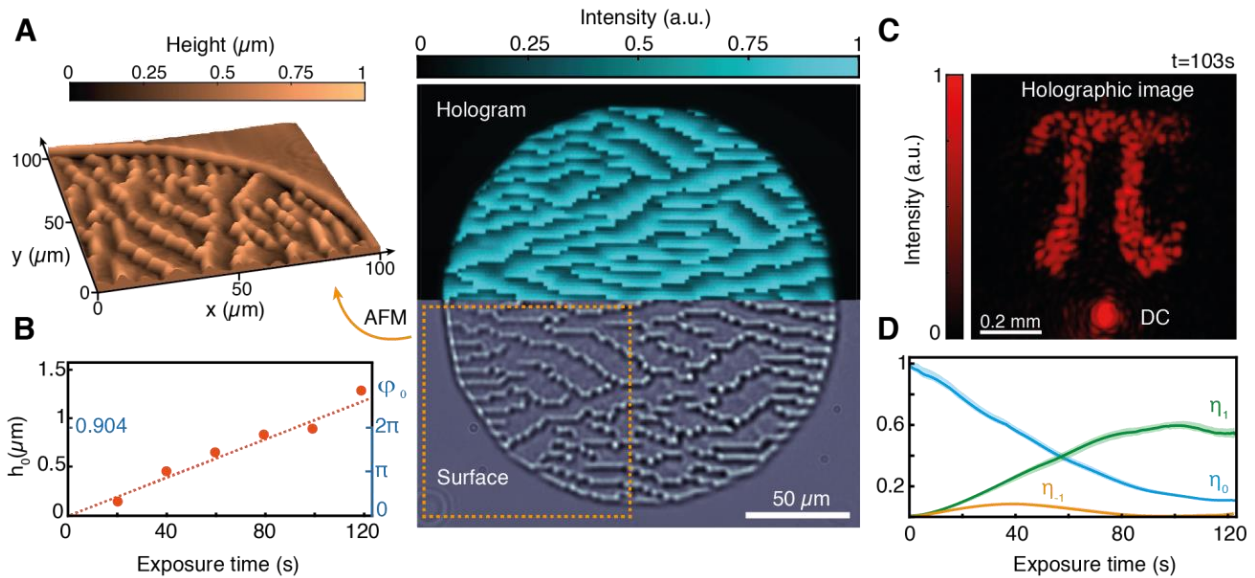

Fig. 3. Fabrication and optimization of azopolymer holographic projectors implemented as kinoforms. The middle panel shows the grayscale holographic pattern reproducing the kinoform design and the resulting SEM image of the structured surface after the exposure. **A** Atomic force microscope (AFM) scan of a quarter portion of the structured surface (100 X 100 $\mu m$) collected right after the exposure process. **B** Full modulation depth $h_0$ as function of the total exposure time. Experimental data are fitted with the model trend $h_0 = c \cdot t$, allowing for the experimental determination of the surface inscription efficiency $c = 10.5 \pm 0.5 \, nm/s$. Blue axis shows the implemented phase depth for a probe wavelength $\lambda_P$. **C** Diffraction pattern acquired at the optimal exposure time, maximizing the diffracted light power effectively shaped in the target holographic image. **D** Experimental trend of the diffraction efficiency reconstructed during the inscription process. Trends are the results of five independent exposures: the average value for the experimental diffraction efficiency at each exposure time is represented by a solid line. The shadow represents the punctual standard deviation.

To quantitatively evaluate the quality of the fabricated surface relief pattern with respect to the design, we retrieved the surface height distribution from the AFM analysis. The topographic distribution is then transformed in a phase delay distribution (by eq. S1) and compared with the phase distribution extracted from the designed phase map, (additional details are presented in Fig. S7-9). We used the Root Mean Square Error (RMSE) to quantitatively definethe average mismatch errors occurred during the fabrication step. The analysis, repeated for different exposure times, provided a constant RMSE, ensuring that any topographical mismatch, related to the hologram design and to the material response, is not worsened by increasing the surface modulation depth $h_0(t)$ to reach the target $h_0$. From the height distributions obtained with fixed illumination parameters at different exposure times, we also determined an experimental estimation of the writing efficiency parameter $c$ entering in eq. (1). We extracted the full relief modulation range from the retrieved distributions to estimate the total modulation depth $h_0(t)$, whose experimental results are provided in Fig. 3B. Those results allowed the empirical definition of the exposure time that provides the optimal $2\pi$ modulation depth in the kinoform for the probe light wavelength of $\lambda_p = 632.8\ nm$. A total exposure time of $t = 86\ s$ is sufficient for optimal inscription of the considered kinoform fabrication to in our experimental conditions.

Nevertheless, this off-line structural characterization roadmap does not guarantee a standardization of the manufacturing process. A new calibration step would be necessary for each different relief geometry and illumination parameters, leading to a time consuming and multi-step workflow.

However, the surface relief pattern developing on the azo-resist can be characterized directly during the surface structuration, providing a real-time feedback on the writing process. Despite different techniques based on mechanical (*62*) and optical (*45*) real-time topographic investigation have been successfully proposed, they do not directly characterize the optical performances of the diffractive surface. On the contrary, the all-optical lithographic scheme proposed here easily allows the direct evaluation of the optimized writing parameters from the analysis of the developing holographic diffraction pattern (*46*, *63*), to act also on specific aspects relevant for applications, as the suppression of the ghost holograms.

To this aim, we illuminated the developing morphological holographic plate on the azo-resist film with an additional laser beam at the probe wavelength $\lambda_p$ during the surface writing step. The developing diffraction pattern is continuously recorded with a CCD, at a repetition rate of 5Hz, during the exposure (Fig. 3C). For each of the acquired frames, we evaluated in real-time the relative diffraction efficiencies $\eta_i$ in the target holographic image, and in the spurious terms (DC order and the ghost image) (Fig. S10).

Fig. 3D summarizes the experimental results for five independent kinoform fabrications. The optimal exposure time ($t_{opt} = 103 \pm 1$ s) was chosen such that the light power diffracted in the holographic target image is maximized. In this condition, in experimental efficiency $\eta_{+1} = 0.60 \pm 0.02$ was obtained. We also observed a relative transmissivity ($|t(x,y)|^2$) equal to 0.96 for the final developed surface (Fig. S11), demonstrating also minimal influence of possible unfavorable light scattering sources produced by the lithographic process. Our approach demonstrates the big advantages offered by a single-step and all-optical structuration technique, allowing the tuning of the optimal exposure parameters in real time, which leads to a fully working device right after its inscription without the need of further time-consuming surface analysis or preliminary calibration procedures.

The off-axis hologram design, analyzed here mainly to highlight the characteristics of our holo-lithographic scheme has a fundamental limitation in practice due to the presence of ghost holograms simultaneous to the target holographic image. In every physical device with unavoidable structural mismatches in the kinoform fabrication, this imposes a having for the exploitable holographic plane and a physical filtering process for the spurious terms.

However, in many applications such as augmented reality and wearable holographic projectors, the holographic image could be formed in a very specific plane of the optical axis, which typically coincides with the observer's eye or with a detector sensor (*1*). When appropriately designed and fabricated, a holographic plate operating in this configuration allows to overlook the presence of any other spurious diffraction order, relaxing also eventual design constrains.

An additional advantage of kinoform-based holographic projectors is the possibility to encode multiple optical functionalities in the same substrate, multiplexing, during the design, the optical properties that two or more phase masks would have exhibited individually. Multiplexing has no impact in terms of calculation resources during the design step, and it can easily explored by the unique combination of our material and the holographic setup (*64*).

Starting from the target phase mask, e.g. resulting from kinoform calculation, an additional proper phase mask can be superimposed to produce an axial shift of the target holographic image with respect to the ghost and DC orders (Fig. 4A). This task can be achieved in an equivalent way by making the light passing in an additional lens of focal length $f$, so the kinoform $\varphi(x, y)$ must be multiplexed with the phase shift produced by a thin lens (*10*), equal to $\varphi_L(x,y) = \pi/\lambda f (x^2 + y^2)$. As the phase of the beam after passing through the phase mask is required to be modulo $2\pi$, the resulting multiplexed phase map (*65*) $\varphi_M$, to be converted in the holographic writing pattern, is $\varphi_M = (\varphi + \varphi_L)_{mod(2\pi)}$. Form the Fourier transform relation (eq. (2)) it can be easily demonstrated, using the generalized Fourier analysis (*61*), that each diffraction order $i$ is axially splitted along the optical axis and it is reconstructed in a different plane located at $z = i \cdot \Delta z$, where $z = 0$ denotes the reconstruction plane of the kinoform without the additional lens phase map. The distance $\Delta z$ is function of the focal length $f$, which determines the axial separation between the holographic image and the other (spurious) orders (Fig. 4B).

Fig. 4C shows a SEM image of the surface relief pattern inscribed on the azo-resist surface using such multiplexed kinoform design. The corresponding diffraction pattern in the target reconstruction plane is presented in Fig. 4D. In this plane of the optical axis, only the target holographic image was clearly visible, while the out of focus DC and ghost terms contributed only with negligible background in the image.

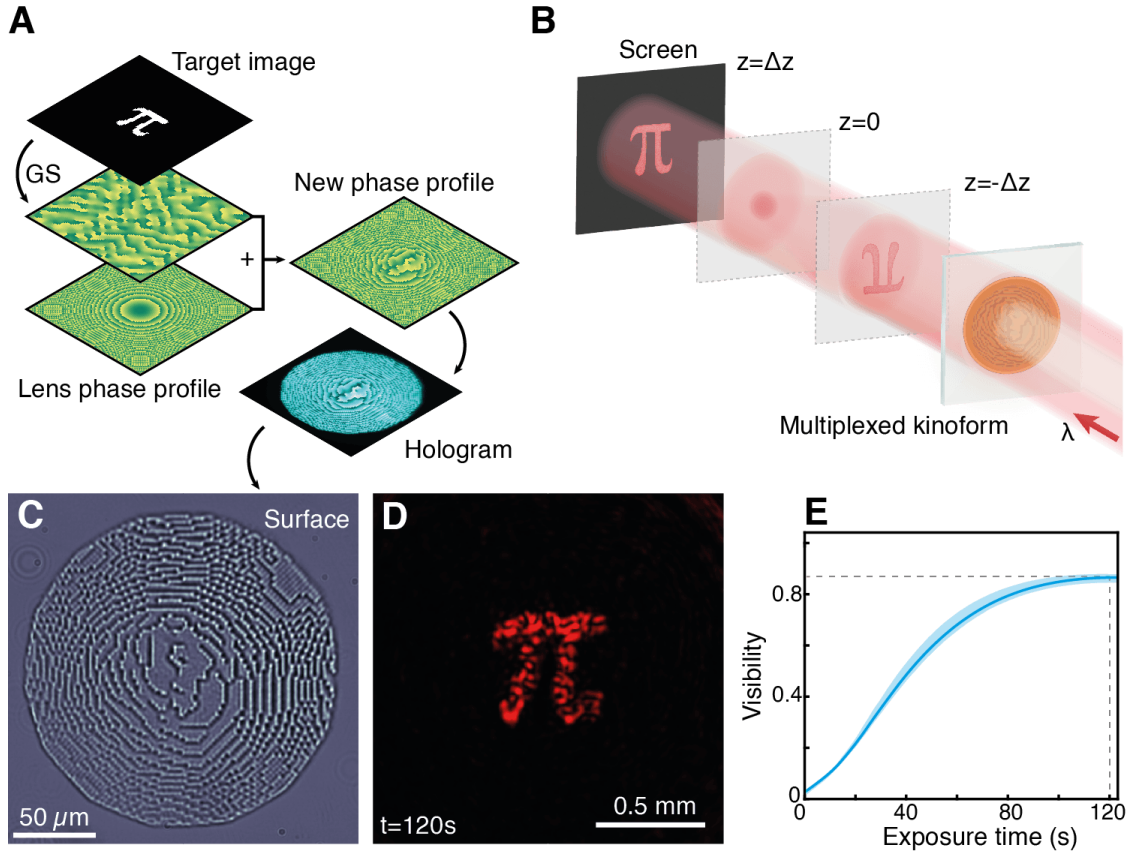

Fig. 4. Design, fabrication, and optimization of multiplexed kinoforms. **A** The resulting kinoform, from a GS algorithm performed on the on-axis image of the letter pi, is multiplexed with a spherical phase profile. The new phase profile is used to encode the different intensity levels of the writing beam. **B** Representation of the diffractive behavior of a multiplexed kinoform. When illuminated with monochromatic coherent light, different diffractive orders are axially reconstructed on shifted planes. Assuming that $z = 0$ is the plane where the holographic pattern is reconstructed without the multiplexing process, each diffraction order $i$ is reconstructed at $z = i \cdot \Delta z$. **C** SEM image of the azopolymer surface after the exposure to the holographic beam for $t_{opt} = 120\ s$. **D** Resulting diffraction pattern acquired at $z = \Delta z$ **E** Experimental trend of the pattern visibility reconstructed during the inscription process as result of five independent exposures: the average value for the pattern visibility at each exposure time is represented by a solid line. The shadow represents the punctual standard deviation.

As we could not simultaneously access to all the diffracted orders during the surface developing to define the relative diffraction efficiency in the holograms, we used the image visibility $\mathcal{V}$ as quality estimator for the light pattern in the target reconstruction plane (additional details are presented in Fig. S12). Similarly to the previous case, the real-time control of this parameter allowed us to directly optimize the exposure time $t_{opt} = 120 \pm 1\ s$ for maximum visibility of $\mathcal{V}_{max} = 0.83 \pm 0.03$ (Fig. 4E). This high contrast image was also the result of an independent tuning of the multiplexed focal length $f$, chosen, according to our setup resolution limit, to maximize orders separation and subsequently the holographic image contrast (Fig. S13).

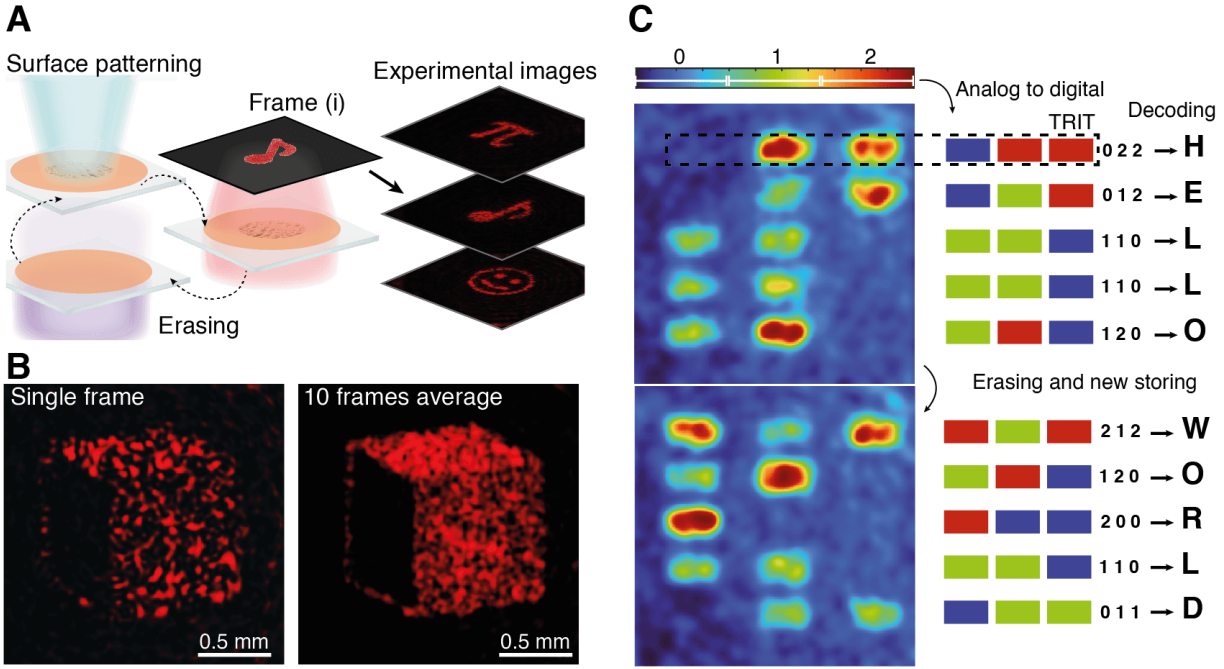

Fig. 5. Fully reprogrammable kinoform for time average image quality improvement and data storing and sharing. **A** Reprogrammable holographic projector: after surface pattering and holographic image acquisition, morphology can be completely restored to pristine flat state, allowing for a new patterning step. Quality enhanced experimental images are the result of the time averaging of multiple holographic patterns. Full resolution images are provided in Fig. S14. **B** Experimental results of holograms time averaging for speckle noise effects reduction. On the left is showed the grayscale pattern acquired after a single exposure step while on the right the same pattern is reconstructed as time average of ten independent exposures over the same azopolymer area. **C** Experimental results of the holographic data storing and sharing. Holographic patterns are plotted with a rainbow colormap. Blue-indigo, green-yellow and orange-red colors are respectively related to three possible intensity levels encoding three digital logic states. Experimental images are converted from an analog to digital map for information readout. Word "HELLO" is reconstructed after a first step of surface writing loop followed by a second multiple exposure step allowing for the reconstruction of the second part of the message, "WORLD".

As additional requisite for the use of morphological holographic projectors in real photonics applications, ranging from optical cryptography to holographic refreshable displays, the surface morphology should be completely reversible and reprogrammable on demand. One of the interesting features of azopolymers is that when illuminated with unstructured light in the chromophore absorption band (see also Fig. S3), the pristine flat surface can be optically restored, allowing multiple and reversible patterning cycles (*46*, *66*). Fig. 5A schematically shows this all-optical reprogrammable surface structing process.

One of the features of dynamic holographic platforms (e.g. LCOS SLMs or DMDs) is that the temporal coordinate can be exploited to produce effective holographic patterns with either enhanced lateral complexity (*64*) or higher image quality (*67*). In these processes, the final holographic image is the result of the temporal average of the individual patterns that are instantaneously produced by the dynamically changing diffractive device. The unique reversible photo-mechanical properties of the azopolymer used here as can be exploited to achieve similar effects. To demonstrate an example practical relevance for our dynamically-evolvign morphological holographic projectors, we repeatedly reprogram the kinoform written on the surface of the azo-resist to produce a time-averaged holographic diffracted image with a reduced speckle noise, intrinsically associated to the kinoform design with a IFT algorithm (*68*, *69*).

Fig S.14 in the Supplementary Information shows the details of the characterization of the holograms recorded in a typical dynamical kinoform reconfiguration experiment. The procedure for the improved average holographic image started by irradiating the pristine azopolymer surface with a holographic writing kinoform (in the multiplexed design). After an inscription process providing optimized visibility in the diffracted holographic pattern, an image $I(x, y, 1)$ of the hologram was collected by the CCD and stored as single frame of a holographic projection movie. At this stage the surface was completely (optically) erased, and the same area of the azo-resist was exposed with a new independently calculated holographic writing pattern, characterized by an independent random distribution of speckle grains. This loop was iterated, acquiring the relative holographic image $I(x, y, i)$ each time. After $N = 10$ writing/erasing steps, the time averaged holographic image was calculated as $\langle I(x,y)\rangle = (N)^{-1}\sum_{i=1}^{N} I(x,y,i)$. As expected, the averaged image is characterized by a speckle severity reduced by a factor $1/\sqrt{N}$, as demonstrated in Fig. S14 for three different target holographic images. This artificial image improvement trough a time averaging process is the same as that performed by an ideal "slow eye or detector", which has a time response much higher than the typical surface reconfiguration time ($\sim 120\,s$ in our experimental condition). Despite still far from the refresh rates achievable with other dynamical systems, these results allow us to include for the first versatile dynamical modulation capabilities for applications with a planar optical diffractive component.

As additional proof, we show the speckle noise time filtering for a three-level target holographic image. Fig. 5B shows the diffraction pattern $I(x, y)$ and the corresponding time averaged holographic pattern $\langle I(x,y)\rangle$ representing the image of a cube, where each of the three displayed faces encode a different diffracted intensity level. The grayscale nature of the hologram became visually clear only once the that the speckle noise contrast reduction is performed (see also Fig. S15), with a significant improvement with respect to a single holographic image.

Morphological reprogrammable devices able to also encode grayscale optical information can represent a valid platform to store encrypted optical information. The use of non-binary bits of light can increase the storage capacity, while simultaneously reducing the required space on the physical support (*70*). We used our azopolymer as a morphological holographic memory support where the visual information was encrypted in the surface topography. The secret message, displaying the word *"HELLO"*, was converted into a ternary base where each letter is codified into three different trit (ternary digit), each assuming three separate logic states. The trits that defines each letter of the word have been arranged in rows to form a three-level grayscale image, where each level corresponds to one of the three possible logic states. The details of the designed ternary alphabet are discussed in Fig. S16. When this image was used to define surface morphology and transferred to the azo-resist surface, all the original information was encrypted by the Fourier transform algorithm, therefore, information readout is possible only optically by means of a proper optical setup (Fig. 5C). Fourier-transform coding also offers the advantage that if part of the surface would be damaged or destroyed, reading the secreted information would still be theoretically possible. We finally completely erased and reshaped the surface geometry to share the second part of the secreted message composed by the word *"WORLD"*. This temporal holographic splitting of the message enhances encryption capabilities and information sharing security. Additionally, it prospects azopolymer structured films as promising reversible high-density memory substrates. We further estimated that by a single surface illumination process, with the defined architecture, we are capable of simultaneously encoding 3,125 bytes of information in a secreted hologram.

## 3. Discussion and Conclusions

Our direct all-optical maskless lithography, using azopolymers as photoresist, represents the state of the art as fabrication technique of fully reversible diffractive flat optical elements with arbitrary holographic pattern reconstruction capabilities. In the simple case of binary modulation for the writing beam demonstrated in this work, we proved our ability to faithfully transfer, in a pure optical process, complex bidimensional geometries as a two-level surface modulation of an azobenzene-containing polymer film. This process, in another perspective, can also be interpreted as a form of information storing if the target image (e.g. the QR code image) is seen as the information to be encrypted as surface morphology on the azopolymeric film. An additional morphological analysis of the surface, right after the exposure process, demonstrated no significant information losses in the morphological information transfer, even considering the differential light response of our material to the writing illumination.

As additional milestone of such method, we extended and scaled our approach for the realization of diffractive kinoforms, where complex lateral geometries with grayscale modulation depth are simultaneously required. The additional possibility to test the devices functionality during the fabrication process provides a cost-effective design and prototyping of operating diffractive optical devices, implemented as azopolymer phase retarders. We characterized both the surface morphology and its diffractive behavior right during the exposure, investigating the quantization and pixelation effects and non-linear responses of the material to the structuring technique, enhancing their relevant impact during the device optimization and fabrication. Our approach led to the realization of pixel-free morphological holographic projectors, ensuring high efficiency and ultra-compact devices, whose depth results comparable with the operating light wavelength. The opposite happens in conventional digital devices, where the discrete nature of the pixels limits the spatial resolution and the addressable phase sampling, while simultaneously generating spurious periodic replication of the reconstructed image, with a consequent overall efficiency loss. Despite simple, morphological encoding design of dielectric diffractive surfaces totally changes the perspective when holographic projectors are also compared to traditional wide displays. First, the complex modulation provided by the realized kinoform has an almost unitary transmittance, resulting in a lossless structuring of light. Furthermore, the Fourier relationship linking the modulation and the image reconstruction plane is non-local, meaning that each of the point of the kinoform will contribute to form the entire holographic image. In other words, a kinoform preserves the information content in all its parts, consequently breaking or damaging the device will not compromise at all the holographic image reconstruction.

Additionally, as the azopolymer surface can be optically restored to the flat pristine state in place, multiple writing/erasing cycles can be performed on time scales of few minutes. As, up to now, no material and structuration method combination for such dynamically changing surfaces exist, our approach represent the state of the art for reversible, all-optical custom flat optical devices fabrication. This possibility allowed time-averaged enhanced quality holographic images and paved the way for the fabrication of morphological reshapable devices able to encode optical information with both morphological and temporal encryption. As valid every time that information needs to be stored on a physical support, the main requests for the substrate are time stability and reversibility. On the other side also the encoding process is required to be highly controlled, as any critical issue may result in information degradation or even in its loss. We demonstrated that azopolymers, when illuminated with digital reconstructed intensity patterns of light, can meet those requirements. For the first time we showed that azopolymer unique optical properties can also be exploited to implement a new class of photonics devices with several applications ranging from wearable holographic projectors and displays to high quality supports for data storing, encryption and sharing. Even if still at a primitive level, this approach already makes evident the benefits that can completely change our prospective for holographic displays, optical data storage, and encryption, opening also to practical applications in emerging technologies as VR\AR displays and wearable devices.

## Materials and Methods

**Experimental setup**

The experimental configuration for the azopolymer surface relief inscription is based on a phase-only Computer-Generated Holograms (CGHs) scheme. Its schematic representation is shown in Fig. S2. A laser diode source (Cobolt Calypso) produces a TEM00 beam at wavelength $\lambda$=491 nm which, after a beam expander (lenses $L_1$ and $L_2$), is phase-modulated by a computer-controlled reflective phase-only Spatial Light Modulator (SLM, Holoeye Pluto). The modulated beam is propagated through a 4$f$ lenses system with the input plane located in the SLM plane. The output plane coincides with the back focal plane of an infinity-corrected long-working distance 50X objective (Mitutoyo), with numerical aperture NA=0.55. The focal lengths of the lenses $L_3$ (300 mm) and $L_4$ (175 mm) are chosen to maximize the spatial resolution in the hologram reconstruction plane. This choice also defines the diameter (~200 μm) of the accessible circular area in the objective front focal plane, which can be used to structure the azopolymer surface in a single illumination step. The position of the sample near the objective focal region is accurately controlled by means of a x-y-z translation stage. Average intensity in the range 12.7-14.0 W/cm$^2$ and circular polarization are used for the structuration of the azopolymer surface. To reduce the speckle noise contrast effects (*67*), the holographic illumination over the azopolymer surface is the result of the time average of several holographic patterns generated from different kinoforms. Each pattern is reconstructed after an independent design from the same target image, initializing the algorithm with random phase. The SLM refresh time (30 Hz for this work) is faster than the azopolymer response so that the effective illumination profile is the temporal average of the illumination profile associated with each of the many independent kinoforms sent in sequence to the modulator. For visual inspection, and proper focusing of the holographic pattern on the photoresponsive surface, a 70/30 beam splitter, placed in the light-path, redirects the light retroreflected by the surface and re-collimated through the objective toward a tube lens (with focal length equal to 200 mm). This lens forms an image of the holographic pattern in its second focal plane, where a "DCC3240M Thorlabs" CCD camera is positioned. During the exposure, an additional diode laser beam at 405 nm illuminates the photoresist film from the substrate side. The beam has circular polarization and different intensity levels depending on its intended function. When the intensity is 0.6 W/cm$^2$, the beam favors the surface structuring process, acting as a writing assisting beam. At intensity higher than 0.9 W/cm$^2$, its absorption causes the erasure of previously inscribed surface structures, acting as an erasing beam. Further characterizations about assisting/erasing beam are described in a previous work (*46*). An additional He-Ne laser beam, at 632.8 nm, is used as sample back-illumination source to test diffraction behavior of the modulated surface during the structuration process. The beam splitter also allows the collection of part of this light without interfering with the writing process, Fig. S3. The image of the surface is projected on the back focal plane of the tube lens and coupled by means of a mirror (mounted on a flip mount) to an additional 2f system composed by the lens L$_5$ (300 mm). Fourier transform image is captured with an additional CCD camera.

**Azo-resist synthesis**

The photoresponsive material used in this work is an azobenzene-containing polymer (azopolymer) in amorphous state, Fig. S3. All reagents were purchased from Merck and used without further purification. The azopolymer was synthesized, purified and characterized as previously reported (Mw = 27000; phase sequence: Glass 67 °C Nematic 113 °C Isotropic; λmax = 350 nm) (*46*, *54*, *71*). The solution for film deposition was prepared by dissolving 70 mg of the polymer in 0.50 ml of 1,1,2,2-tetrachloroethane and filtered on 0.2 µm PTFE membrane filters. The desired film thickness (typically $1.5 \pm 0.1\ \mu m$) was obtained by spin coating the solution on 24x60 mm cover slides at 300 rpm for 4 minutes. In the final stage, the samples were kept under vacuum at room temperature for 24 h to remove solvent traces. Molecular structural formula and the absorbance in the UV-visible are provided in Fig. S3.

**Morphological characterization of structured surfaces**

Topographic characterization of inscribed azopolymer surface reliefs is performed using AFM and SEM. For AFM measurements, a WITec Alpha RS300 microscope is used. The AFM is operated in tapping mode using a cantilever with 75 kHz resonance frequency and nominal force constant of 2.8 N/m. AFM tips (Arrow FM type from Nano World), with nominal radius of curvature of ≈10 nm, are used. The maximum scanned area has a size of 100 × 100 µm$^2$, acquired with resolution of 500 points per lines and 500 lines per scan. For each AFM the minimum of the topography is set to zero to extract the height distribution $P_j$, representing the probability to find a pixel in the image with a height value between $h_j$ and $h_{j+1}$ where $h_j = j\Delta h$. Here $j$ ranges from zero to $N-1$ where $N$ is the number of occupied bins in each image, while $\Delta h = 10 nm$ represents a reasonably choice for the fixed bin width. Each height distribution is normalized to match the condition $\sum_N P_i = 1$. The expected value $\bar{h} = \sum_N h_i P_i$ and variance $\sigma^2 = \sum_N (h_i - \bar{h})^2 P_i$ are extracted for each distribution. To retrieve an estimation of the modulation depth $h_0$ we consider the discrete integral function $I(n) = \sum_{i=0}^{n} P_i$. Firstly, we define $h_0 = h_k$ where $k$ satisfies the relation $I(k) = 1$. In that case we represent the total modulation depth as the full dispersion range of the distribution $P_i$. Since, due to our material behavior, the height distribution is not uniform, we also estimate the modulation depth as $h_0 = h_a - h_b$ where $a$ and $b$ satisfy respectively $I(a) = 0.95$ and $I(b) = 0.05$; with this assumption $h_0$ represents the range, uniformly distributed around the median of the distribution, where there is the 0.90 of probability to find a fixed value of $h_i$, see Fig. S7.

Scanning electron microscopy (SEM) images are acquired with a field emission gun (FEG–SEM) FEI/ThermoFisher Nova NanoSEM 450 microscope. Samples are sputtered with a layer of Au/Pd using a Denton Vacuum Desk V TSC coating system prior to observation.

**Iterative Fourier transform algorithms**

Despite the simple Fourier relation, optical modulation cannot be retrieved by simply inverting the equation (2). To design the proper phase mask able to lossless transform a given input light distribution into a desired light pattern, an iterative Fourier transform algorithm (IFTA) has been used. In this class of algorithms, the optical field is bounced back and forth between two planes related by a Fourier transform, applying specific constraints to the retrieved fields at each iteration. The used algorithm for diffractive kinoforms design is the Gerchberg-Saxton algorithm (*72*). This algorithm can be easily implemented with modern computing capabilities, and once a digital representation of $I_{out}$ is provided by a grayscale 8-bit digital image it returns a digital representation of the phase map $\varphi(x, y)$. When the desired light distribution is only constrained in a limited region of space, as for the holographic writing beam, the complex amplitude outside this area can be arbitrary chosen or left free to vary, allowing to increase the light hologram quality. This possibility is typically referred to as amplitude freedom. We used this approach to generate the writing holograms for the azopolymer structuration by a mixed region amplitude freedom (MRAF) algorithm (*56*). We implemented both GS and MRAF algorithms in MATLAB, using the Fast Fourier Transform (FFT) algorithm.

Supplementary Material for

# Reprogrammable holograms from maskless surface photo-morphing

**Francesco Reda,[1] Marcella Salvatore,[1,2] Marco Astarita,[3] Fabio Borbone[4] and Stefano L. Oscurato[1,2,*]**

[1] *Physics Department "E. Pancini", University of Naples "Federico II", Complesso Universitario di Monte Sant'Angelo, via Cinthia 21, 80126, Naples, Italy*

[2] *Centro Servizi Metrologici e tecnologici Avanzati (CeSMA), University of Naples "Federico II", Complesso Universitario di Monte Sant'Angelo, Via Cintia 21, 80126, Naples, Italy*

[3] *Physics Department, Politecnico di Milano, 20133, Milan, Italy*

[4] *Department of Chemical Sciences, University of Naples "Federico II", Complesso Universitario di Monte Sant'Angelo, Via Cintia, 80126 Naples, Italy*

*\*Corresponding author: stefanoluigi.oscurato@unina.it*

**Diffraction properties of holographic morphological projectors.**

Phase-only holographic plates are typically represented by a complex transmission function $t(x, y)$ describing in the scalar approximation of wave optics, the wavefront phase modulation of an incident monochromatic optical field, at wavelength $\lambda$ passing through the device. Isotropic dielectric phase retarders implement phase modulation as result of local variations in thickness $h(x, y)$ and refractive index $n(x, y, \lambda)$ of the device, whereby the planar modulation function can be written as:

$$t(x, y) = \exp\left[i\varphi(x, y)\right] = \exp\left[i\frac{2\pi}{\lambda}(n(x, y, \lambda) - n_s)h(x, y)\right] \qquad (S1)$$

representing the local phase delay $\varphi(x, y)$ accumulated by the light due to optical path variation imposed by the plate immersed in a surrounding material whose refractive index is $n_s$. When a kinoform, supposed to be at $z = 0$, is illuminated by the light field $U_{in}(x, y, 0)$, the resulting complex field $U_{out}(x, y, z)$ formed due to Fraunhofer ($z \gg 0$) diffraction is the two-dimensional spatial Fourier transform of the modulated beam at the kinoform plane, and the reconstructed image $I_{out}$ is determined by the relation:

$$I_{out}(x, y, z) = |FT[U_{in}(x, y, 0) \cdot t(x, y, 0)]|^2 \tag{S2}$$

The phase encoding process, from the phase design to its implementation by lithography, leads to a device whose real transmittance $t^{real}$, is a function $g$ of the designed phase $t^{real}(x, y) = e^{ig[\varphi(x,y)]}$. To consider any possible deviation from this ideal case, the complex transmittance of the real kinoform can be decomposed into a linear superposition of functions, clearly describing the effects of phase mismatches. Assuming that the deformation is space invariant, the spatial coordinates can be omitted and the function $t^{real}$ can be expanded in terms of its argument, according to the generalized harmonic analysis (61):

$$t^{real} = \sum_{\alpha=-\infty}^{+\infty} G_\alpha e^{i\alpha\varphi} \tag{S3}$$

where $G_\alpha = \int_0^{2\pi} t^{real} e^{i\alpha\varphi} \, d\varphi$ and $\alpha$ an integer. The $\alpha = 1$ term is the only one whose Fourier transform results in an optical field with intensity $I_{out}$. The amount of optical power shaped in the reconstructed intensity profile with respect to the total transmitted power, is equal to $\eta_1 = |G_1|^2$ and it is equal to one only in the ideal case $g(\varphi) = \varphi$. The other terms of the series, apart from the term $\alpha = 0$ which determines an unmodulated optical component named DC term, contribute with shifted and scaled replicas of the desired intensity pattern, known as ghosts or false images (8, 60). The total reconstructed pattern is a weighted sum of the desired image, the DC term, and false images, typically also spatially overlapped in the reconstruction plane and with a relative efficiency $\eta_\alpha = |G_\alpha|^2$. At best, once the geometry $h(x, y)$ is fixed, $g$ is linear with the total surface reliefs amplitude $h_0$, which has to be tuned in order to reach a fully $2\pi$ modulation depth. According to equation (S1) this condition is achieved for $h_0 = \lambda/(n(\lambda) - n_S)$; for our material at the operating wavelength $\lambda = 0.6328 \, \mu m$, $n(\lambda) = 1.696$ and $n_S = 1$ (for air immersed kinoforms), the condition is satisfied for $h_0 = 0.9092 \, \mu m$. In this simple case if $h_0^*$ is the implemented modulation depth, the ratio $m = h_0^*/h_0$ denotes a mismatch parameter. Under those conditions diffraction efficiency $\eta_\alpha$ can be written as (60):

$$\eta_\alpha = sinc^2(m - \alpha) \tag{S4}$$

The condition $m = 1$ guarantees the maximum diffraction efficiency $\eta_1 = 1$, ensuring that all the incident optical power is effectively shaped in the reconstructed holographic pattern [Fig. S1]. During the encoding process, more complex distortion effects can determine a non-linear form for $g$. These certainly include quantization and pixelation effects and non-linear responses of the material to the structuring process that led to a kinoform in which phase mismatches are included providing $\eta_1 \neq 1$ even if target modulation depth is reached.

# Supplementary figures

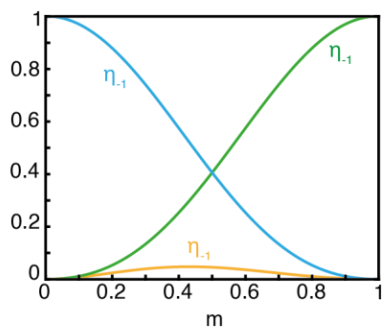

Fig. S1: Theoretical diffraction efficiency from an ideal kinoform as function of the height mismatch error *m*.

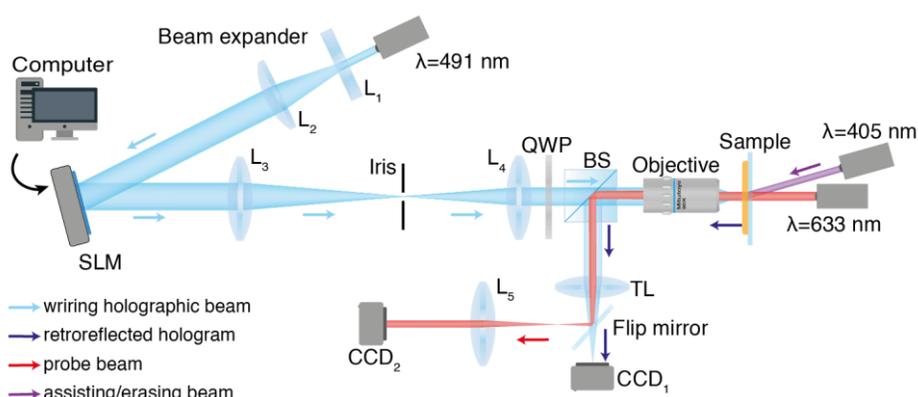

Fig. S2: Schematic of the experimental setup. Beam expander - lenses $L_1$ ($f_1$=-50 mm) and $L_2$ ($f_2$=250 mm). SLM - Holoeye Pluto, LCOS spatial light modulator, phase only (reflective). 4$f$ configuration - lenses $L_3$, ($f_3$=300 mm) and $L_4$ ($f_2$=175 mm). QWP - quarter wave plate. BS - 70/30 beam splitter. Objective - 50X Mitutoyo Plan Apo Infinity Corrected Long WD Objective. TL - tube lens ($f_{TL}$=200 mm). Fourier transforming lens - lens $L_5$, ($f_5$=300 mm). CCD$_{1/2}$ - "DCC3240M Thorlabs" camera.

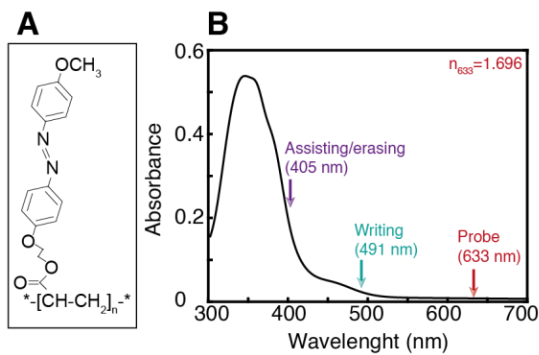

Fig. S3: Azopolymer optical characterization. **A** Molecular structural formula. **B** Absorbance in the UV-visible. Probe wavelength $\lambda_p = 633\ nm$ is chosen different from the writing beam and out of the absorption band of the material. Refractive index at $\lambda_p$ was measured via ellipsometry.

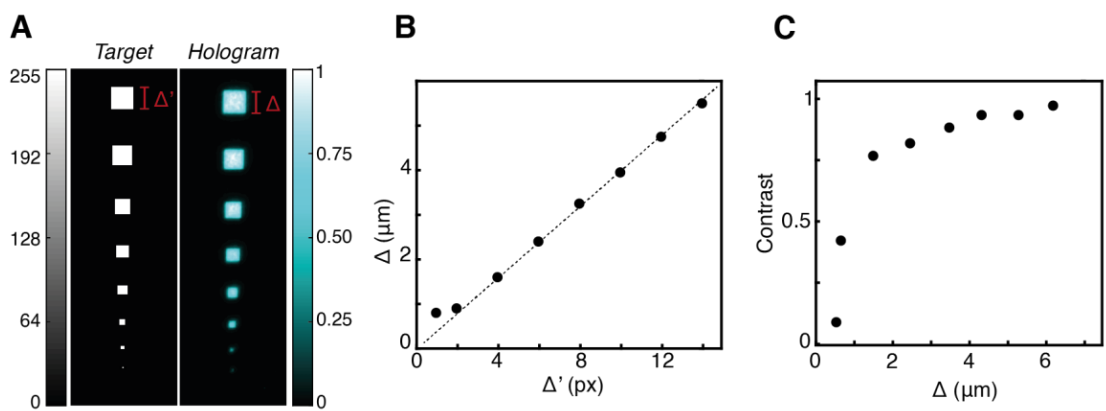

Fig. S4: Holographic setup spatial resolution. **A** Holographic reconstruction of square shaped light pattern with lateral size $\Delta$. **B** Experimental squares size $\Delta$ as function of designed size $\Delta'$. The slope $b = 0.376 \pm 0.002\ \mu m$ of the fitted line trend $\Delta = b\Delta'$ defines the calibration of physical dimensions of patterns in the polymer plane with respect to the analytically designed target images. **C** Contrast of the holographic reconstructed square as function of lateral size $\Delta$. Contrast is defined as $C = (I_W + I_B)/(I_W - I_B)$ with $I_W$ and $I_B$ representing the average experimental intensity levels corresponding to white and black areas of the target image, respectively. Resolution limit is reasonably set to $\Delta_0 = 5b = 1.88\ \mu m$. Kinoforms resulting from design step are scaled by a factor 5 before being encoded into the illumination pattern, ensuring the highest contrast for the holographic writing pattern maintaining a reasonably reduced pixel size.

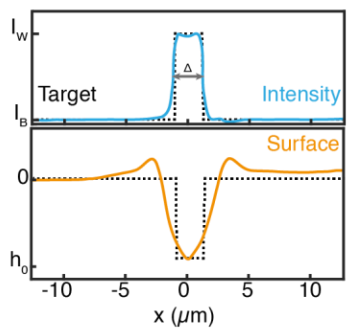

Fig. S5: Azopolymer response to an intensity structured field with designed lateral size $\Delta_0 = 1.88\ \mu m$.

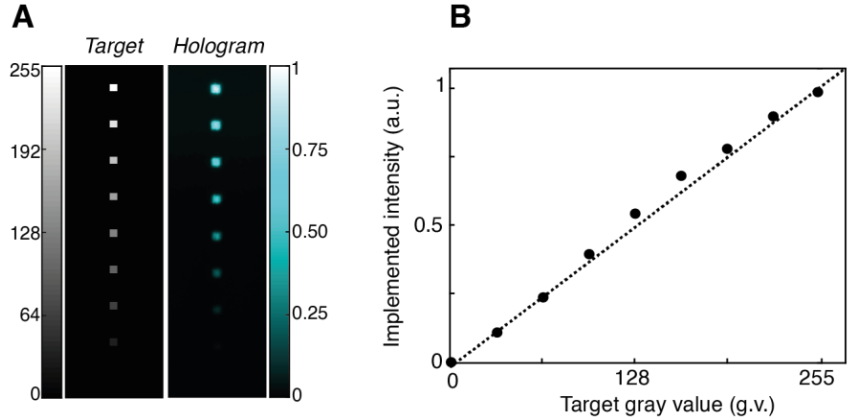

Fig. S6: Holographic setup intensity level modulation. **A** Holographic reconstruction of square shaped light pattern with lateral size $\Delta_0 = 1.88\,\mu m$ and linearly spaced gray levels. **B** Implemented intensity levels as function of the addressed gray value in the target image. The line trend has a slope equal to $0.077\,(a.u.)$.

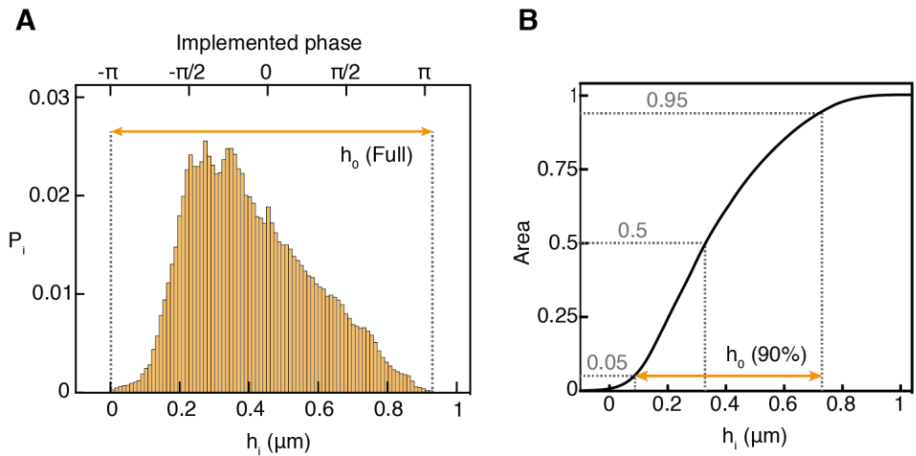

Fig. S7: Height distribution and height modulation depth estimation. **A** Height distribution $P_i$ related to the AFM micrograph presented in Fig. 3A. Full range dispersion, allowing for the experimental estimation of the total modulation depth $h_0 = 0.910\,\mu m$ is defined as $h_0 = N \cdot \Delta h$, where $N = 91$ is the number of occupied bins while $\Delta h = 0.01\,\mu m$ is the fixed bin width. Implemented phase depth is considered for a probe wavelength $\lambda_P = 0.6328\,\mu m$ assuming a refractive index equal to $n = 1.696$. **B** Integral function $I(n) = \sum_{i=0}^{n} P_i$. Experimental estimation of $h_0(90\%)$ represents the height range, uniformly distributed around the median of the distribution, where there is the 0.90 of probability to find a corresponding height value in the AFM micrograph.

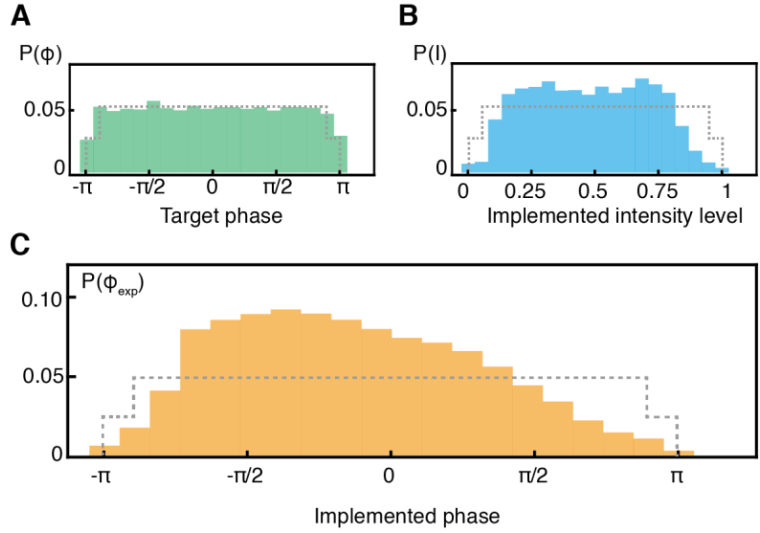

Fig. S8: Comparison between: **A** phase distribution probability in the target phase map resulting from GS algorithm, **B** intensity distribution probability in the holographic pattern and **C** implemented phase distribution retrieved from the AFM image. Implemented phase depth is considered for a probe wavelength $\lambda_P = 0.6328\ \mu m$ assuming a refractive index equal to $n = 1.696$. For visual clarity, data are binned considering $N = 20$.

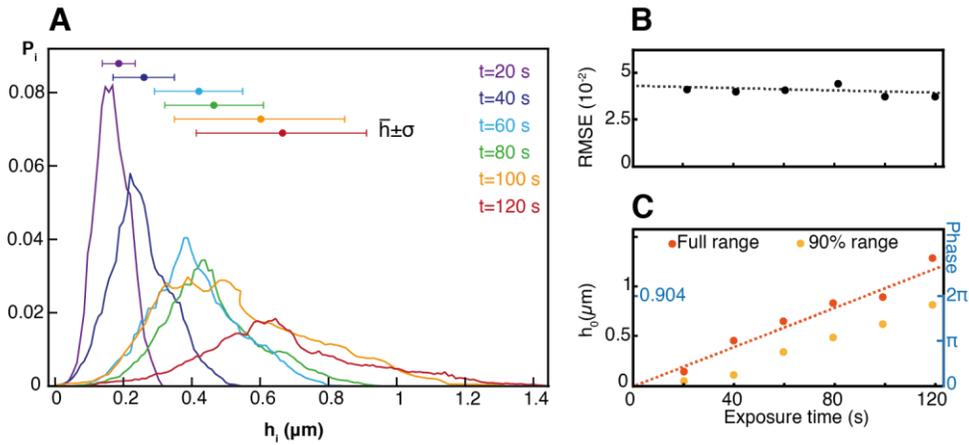

Fig. S9: Temporal characterization of structured surface. **A** Height distribution for six different exposure times. Each dot represents the expected value $\bar{h}$ and relative variance $\sigma^2$ for the corresponding distribution. **B** Root mean square error defined as $\text{RMSE} = \sqrt{\sum_N (P_i - \bar{P})^2}$ as function of the total exposure time. $\bar{P}$ represents the target uniform distribution expected at different exposure times. **C** Full range modulation depth $h_0$ and 90% dispersion range as function of the total exposure time. Implemented phase depth is considered for a probe wavelength $\lambda_P = 0.6328\ \mu m$ assuming a refractive index equal to $n = 1.696$.

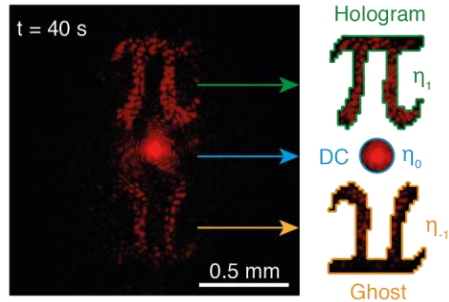

Fig. S10: Experimental determination of diffraction efficiency $\eta_i$ determined by integrating the CCD signal over the regions of interest delimited by the colored trace in the image. Green area corresponds to the holographic image efficiency while light blue and orange area correspond to the DC order and ghost image, respectively.

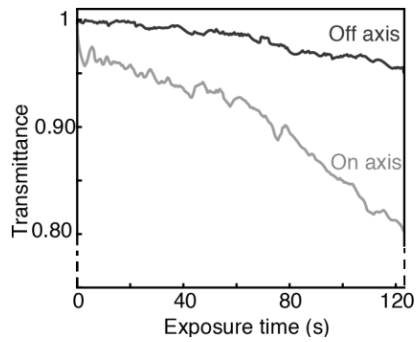

Fig. S11: Kinoform transmittance over exposure time determined by integrating the CCD signal over the full sensor size.

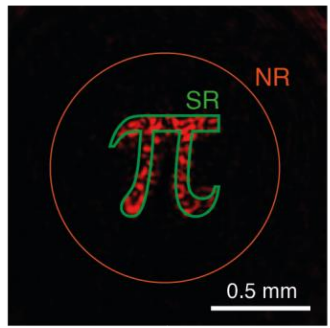

Fig. S12: Experimental determination of pattern visibility. Visibility is defined as $V = (I_{SR} + I_{NR})/(I_{SR} - I_{NR})$ where $I_{SR}$ is the average intensity inside the signal region (green area) and $I_{NR}$ is the average noise level outside the holographic image (orange area).

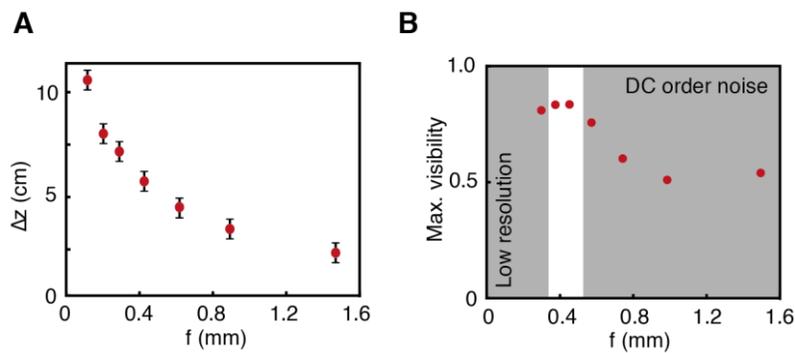

Fig. S13: Optimization of multiplexed spherical profile. **A** Axial shifting $\Delta z$ of the holographic image as function of the spherical phase profile parameter $f$. **B** Maximum visibility achieved with different spherical phase profile parameter $f$. Best value for multiplexed focal length is $f = 0.450\ mm$, allowing for high visibility and reasonable orders separation.

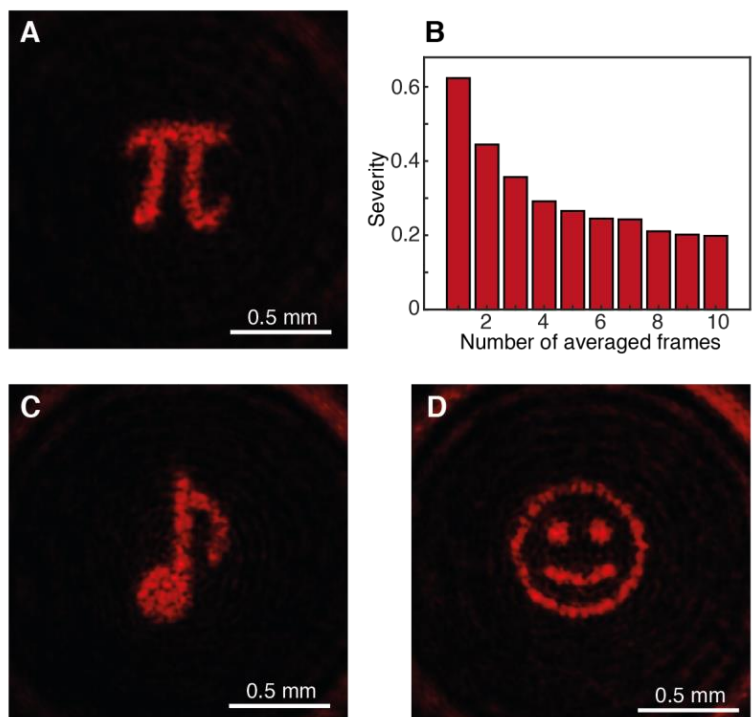

Fig. S14: Speckle contrast reduction process by holograms time averaging. **A** Average holographic pattern acquired after 10 writing/erasing cycles representing the on-axis image of the Greek letter pi. **B** Speckle noise severity as function of the number of averaged frames. Severity is defined as $\sigma/\langle I \rangle$ where $\langle I \rangle$ is the mean intensity and $\sigma$ is its standard deviation measured in the image, see also (*67*). **C-D** Average holographic pattern acquired after 10 writing/erasing cycles representing the on-axis image of a music note and a smile, respectively.

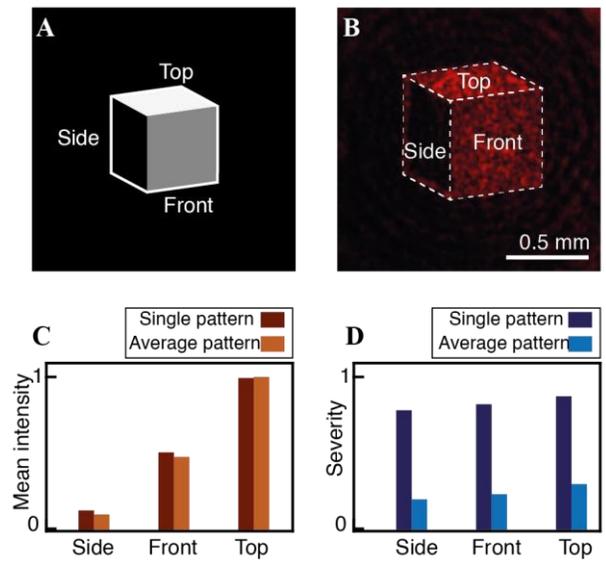

Fig. S15 Speckle analysis of a time averaged grayscale pattern. **A** Target image. **B** Resulting average holographic pattern acquired after 10 writing/erasing cycles. **C** Comparison between the mean intensity level of three cube faces for the single frame and the time average. **D** Comparison between the speckle severity of three cube faces for the single frame and the time average.

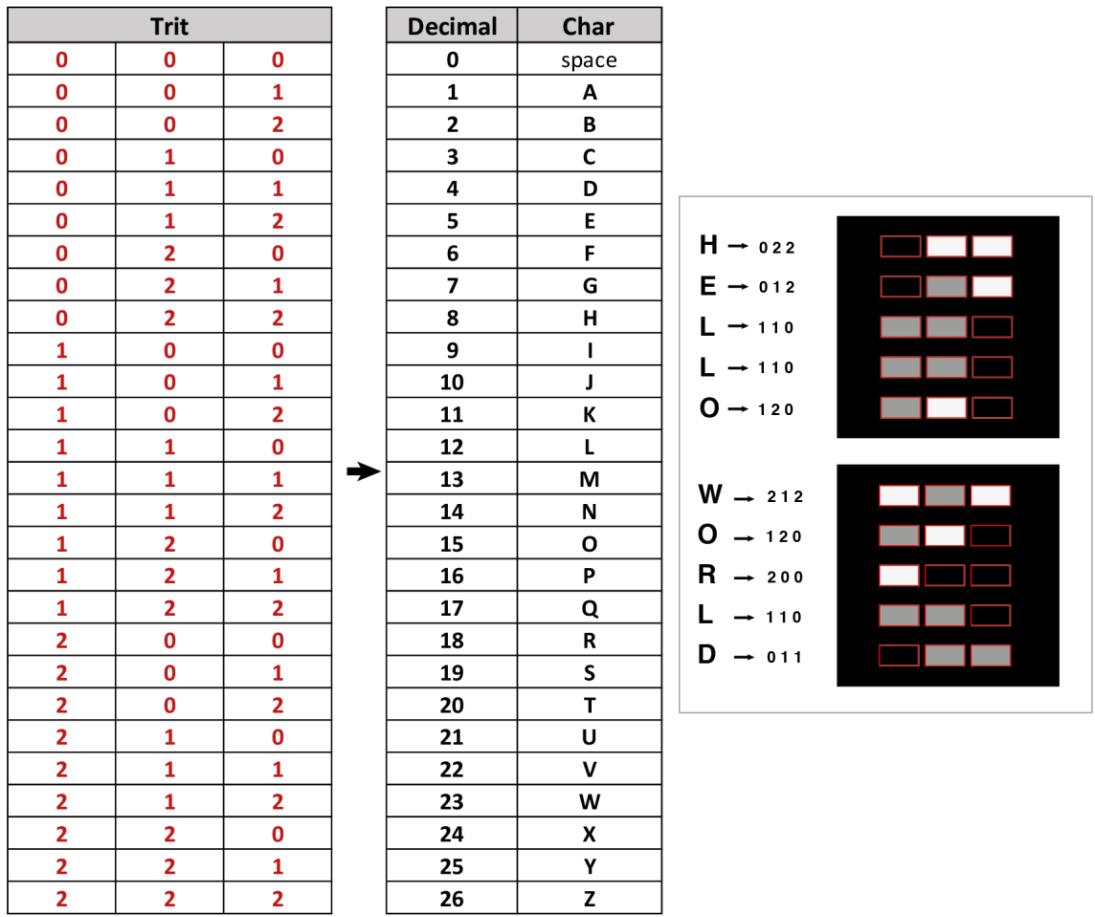

Fig. S16: Look up table for optical encryption and decryption of text messages.